\documentclass[twocolumn]{aastex631}
\usepackage{mathtools}
\usepackage{graphicx}	
\usepackage{amsmath}	
\usepackage{footmisc}
\usepackage{newtxtext,newtxmath}

\newcommand{\code}[1]{{\sc{#1}}}

\usepackage{ulem}
\usepackage{xcolor,tikz}
\newcommand\reduline{%
 \bgroup\markoverwith
  {\textcolor{red}{\pgfsetfillopacity{0.2}\rule[-0.5ex]{2pt}{10pt}\pgfsetfillopacity{1}}%
   \textcolor{red}{\llap{\rule[0.4ex]{2pt}{0.4pt}}\llap{\rule[0.7ex]{2pt}{0.4pt}}}%
  }%
  \ULon}
\begin{document}

\title[Algorithmic Pulsar Timer for Binaries]{Algorithmic Pulsar Timer for Binaries}


\author[0009-0006-8984-9220]{Jackson Taylor}
\affiliation{Department of Physics and Astronomy, West Virginia University, P.O. Box 6315,
Morgantown, WV 26506, USA}
\affiliation{Center for Gravitational Waves and Cosmology, West Virginia University, Chestnut Ridge Research Building, Morgantown, WV 26505, USA}

\author[0000-0001-5799-9714]{Scott Ransom}
\affiliation{National Radio Astronomy Observatory,
520 Edgemont Rd.,
Charlottesville, VA 22903, USA}

\author[0000-0001-5624-4635]{Prajwal V. Padmanabh}
\affiliation{Max Planck Institute for Gravitational Physics (Albert Einstein Institute), D-30167 Hannover, Germany}
\affiliation{Leibniz Universit\"{a}t Hannover, D-30167 Hannover, Germany}




\begin{abstract}
Pulsar timing is a powerful tool that, by accounting for every rotation of a pulsar, precisely measures the spin frequency, spin frequency derivatives, astrometric position, binary parameters when applicable, properties of the ISM, and potentially general relativistic effects.
Typically, this process demands fairly stringent scheduling requirements for monitoring observations as well as deep domain knowledge to ``phase connect'' the timing data. We present an algorithm that automates the pulsar timing process for binary pulsars, whose timing solutions have an additional level of complexity, although the algorithm works for isolated pulsars as well. Using the statistical F-test and the quadratic dependence of the reduced $\chi^2$ near a minimum, the global rotation count of a pulsar can be determined efficiently and systematically. We have used our algorithm to establish timing solutions for two newly discovered binary pulsars, PSRs~J1748$-$2446aq and J1748$-$2446at, in the globular cluster Terzan 5, using $\sim$ 70 Green Bank Telescope observations from the last 13 years.
\end{abstract}



\section{Introduction}

Pulsar timing is the process of monitoring and tracking pulses and times of arrival (TOAs). 
Requiring that each TOA unambiguously corresponds to an integer rotation of the pulsar has made the study of pulsars remarkably precise and quite fruitful since their discovery in 1967 \citep{hewish1968}. Binary millisecond pulsars (MSPs) in particular have allowed pulsar astronomers to validate general relativity \citep{einstein1915} at the $1.3 \times 10^{-4}$ level in the strong field regime \citep{kramer2021strong}. Timing these binary pulsars can also constrain the mass of the neutron star, ruling out several equations of state \citep[e.g.][]{demorest2010two, antoniadis2013massive, cromartie2020relativistic}. 
The International Pulsar Timing Array \citep[IPTA;][]{hobbs2010international} also places heavy importance on timing MSPs. The IPTA is a consortium of consortia with the intention of detecting nHz-frequency gravitational waves (GWs) by correlating the data from around 100 MSPs, most of which are in binary systems. Recent evidence for the detection of these GWs \citep[e.g.][]{agazie2023nanograv, antoniadis2023second, reardon2023search} has solidified the value in timing MSPs for years to come. 

Early on in the pulsar timing process, the global rotation count, or pulse count, from the available data must be determined.
When every TOA is assigned the correct pulse count, this is referred to as the \emph{solution}. In some cases, the data is too sparse to obtain a solution manually. 
With the increasing demand for telescope time, it becomes imperative to develop techniques for achieving a phase-connected solution with a minimal amount of data.
Thus, the Algorithmic Pulsar Timer (\code{APT}) was developed by \citet[hereafter PR22]{Phillips2022} to solve single system pulsars. However, \code{APT} is only successful if the best available model incorrectly predicts the pulse numbers by no more than roughly one or two rotations between each pair of observations. Additionally, this timer falls short of a necessary tool for solving binary systems, that is, JUMPs. A JUMP is a statement applied to a group of several TOAs that allows the fitting software to fit for an arbitrary time offset for that group. 
Another pulsar timing algorithm, developed by \citet[hereafter FR18]{Freire2018} called {\tt DRACULA},\footnote{\url{https://github.com/pfreire163/Dracula}} has seen success. Their algorithm uses JUMPs and works on binaries, but was not explicitly designed for binaries. The fact that {\tt DRACULA} is not fully automated can indeed present a problem for phase-connecting binary pulsars more so than for isolated pulsars due to the need to account for binary parameters. {\tt DRACULA} also does not directly incorporate a procedure for when to fit for additional parameters.

We have developed the Algorithmic Pulsar Timer for Binaries (\code{APTB})\footnote{\code{APT}, \code{APTB}, and \code{APTB}'s User Guide can be found at \url{https://github.com/Jackson-D-Taylor/APT}} to address the shortcomings of \code{APT} and {\tt DRACULA}.
As \code{APTB} was explicitly designed for binaries, it can use any of three different and well-established binary models automatically, and additional binary models can easily be implemented. 
Our algorithm uses the F-test for model comparison so that a parameter is only fit for when its omission becomes significant.
For example, this avoids fitting for right ascension before it is no longer covariant with the spin frequency derivatives.
We also implemented a straightforward tree structure to not only allow \code{APTB} to pursue several options (Section \ref{sec:tree}), but also to easily identify how exactly \code{APTB} phase-connected or failed to phase-connect the pulsar.
Lastly, \code{APTB} is written in Python which permits ease of readability and has the goal of being used by a wider audience. 
As such, we created a user guide
for those who only need to use \code{APTB} for pulsar timing rather than an explanation of our methods. 
It should be noted that the infrastructure required to time binary pulsars is more than sufficient to allow \code{APTB} to time isolated pulsars, as well.
By combining elements of the original \code{APT} and {\tt DRACULA}, as well as our own additions, \code{APTB} explores the best pulse counts in order to potentially yield a phase-connected fit. 
Despite the value of timing binary pulsars, the difficult process is not sufficiently documented in the literature. Thus, an important aspect of this paper is to document one such binary timing method. 

\section{Techniques and Methods}
\label{sec:tech}

\begin{figure}
	\includegraphics[width=\columnwidth]{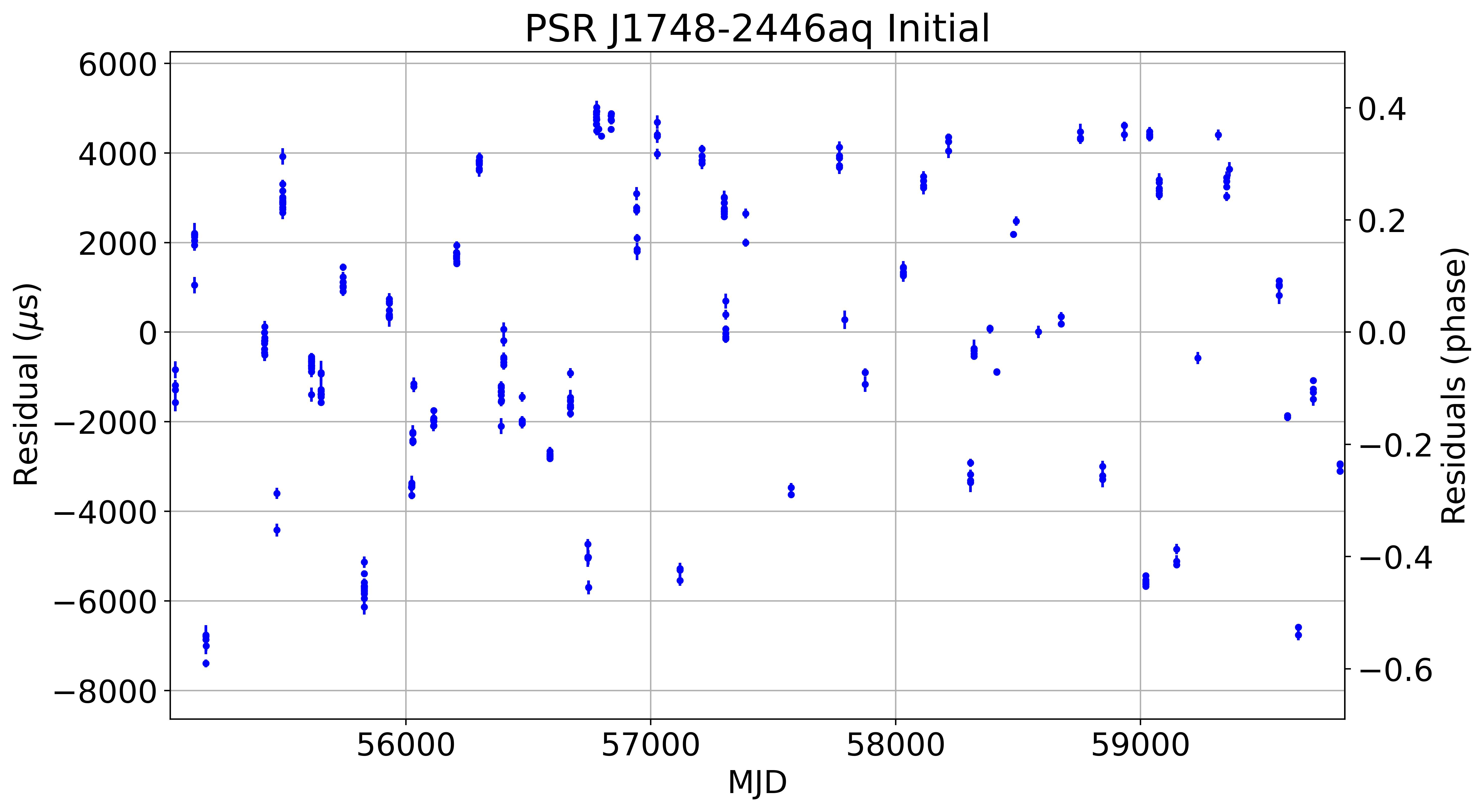}
    \caption{Residuals of the initial timing ephemeris of PSR~J1748$-$2446aq. Notice how the residuals seem to lack any long-term trend and vary greatly from zero. This is in contrast to Figure \ref{fig:psrter5aq_final} where the residuals are nearly zero with respect to their estimated error.}
    \label{fig:wrong_solution}
\end{figure}

Residuals are the difference in phase between the model-predicted TOAs and the observed TOAs. Residuals are modulo the pulse period because initially, we cannot claim to know the global rotation count of every TOA.
It is usually immediately clear when an initial timing ephemeris does not correctly assign the correct global rotation counts because the (incorrect) solution is characterized by seemingly random residuals versus pulse phase (e.g.~Figure \ref{fig:wrong_solution}). Moreover, the initial timing ephemeris holds little predictive power. However, the initial model should be able to ensure that the relative rotation count within a day is correct, i.e. the TOAs in an observation are \emph{phase connected}. TOAs in a single observation belong to a \emph{cluster}. Given the high-dimensional parameter space involved with timing binaries, and multiple important timescales (e.g.~orbital period for binary parameters, annual modulations from astrometric position fitting, and longer timescales, potentially, for spin-down effects), it is necessary to use the information from every TOA, especially those that are known to be phase-connected to other nearby TOAs, even if TOAs between observations are not phase connected. 
By using JUMPs (Section \ref{sec:JUMPs}), the 
residuals within an observation can be minimized with respect to the model parameters.

This improves the model, especially the binary parameters. The process of removing the JUMPs successively is known as \emph{mapping a gap} and is described in Section \ref{sec:gapmap} of this paper as well as Section 4.1 of FR18. All fitting and analysis in \code{APTB} is done via the \code{PINT} pulsar timing package \citep{luo2019, luo2021pint}.\footnote{\url{https://github.com/nanograv/PINT}}

\subsection{JUMPs}
\label{sec:JUMPs} 
Even though the initial timing ephemeris gives residuals that appear random, there are in fact patterns within each cluster (i.e.~single observation or closely spaced sets of observations). \code{PINT} has the infrastructure to extract valuable information from these patterns by fitting each cluster for a different arbitrary time offset. More specifically, \code{PINT} will find the parameters, including time offsets, such that the \emph{reduced} $\chi^2$ ($\chi^2_\nu$, where $\nu$ represents the degrees of freedom) of the model is minimized. Allowing \code{PINT} to fit a cluster for an arbitrary time offset is called applying a \emph{JUMP} to that cluster. The arbitrary time offset itself may be called a JUMP as well.
The reason this method obtains a better solution is that the JUMPs account for unknown phase counts between clusters while the main timing parameters change to account for the phase behavior within each and every cluster. This is particularly useful for finding accurate binary parameters as these can have large effects on short timescales. When JUMPs are applied to multiple clusters, information on the pulsar phase between the clusters is lost. Thus, the JUMPs have to be sequentially removed to obtain a full solution.

To initialize a model, each cluster except one should receive a separate JUMP. At this point, fitting the TOAs for several parameters, like spin frequency ($f$ or $F0$) and a few binary parameters should give a very low $\chi^2_\nu$. This $\chi^2_\nu$ should be very close to unity, and a $\chi^2_{\nu} \gtrsim 2$ should be cause for alarm---the initial ephemeris is likely too erroneous (e.g.~an inaccurate estimate of the initial binary parameters) or the TOA errors have been underestimated. The initial $\chi^2_\nu$ will be referred to as the base $\chi^2_\nu$, or $\chi^2_{\nu,\text{ base}}$. 

\subsection{TOA Scoring}
\label{sec:scoring}
In this section, we will describe why it is important to rank clusters in terms of importance, and the way that we rank them. When initializing a model, not every cluster can be JUMPed---there needs to be an anchor cluster. This first non-JUMPed cluster will be referred to as the \emph{starting cluster}. As described in Section \ref{sec:algorithm}, the starting cluster can have a large impact on whether \code{APTB} can discover a solution. It is thus prudent to find the best starting cluster before actually attempting any solution, as running \code{APTB} for each and every possible starting cluster would be computationally expensive and likely unnecessary. 

We thus adopt a scoring system discussed in Section 4 and Equation~2 of PR22, with only a minor change. This scheme scores each TOA based on how close it is to all other TOAs. A dense group of TOAs will generally all have high scores, while isolated TOAs will ideally be given low scores. Our change to the method described in PR22 is that we apply an index, $\alpha$, to the weighting scheme such that the score of the $i^{th}$ TOA, $n_i$,  reads
\begin{equation}
    n_i = \sum_{j\neq i}\frac{1}{\lvert t_i - t_j \rvert^\alpha},
	\label{eq:score}
\end{equation}
where $t_j$ refers to the arrival time of the $j^{th}$ TOA and $t_i$ refers to the arrival time of the TOA being scored.
A cluster's score is the highest score of its constituent TOAs.

\code{APTB} currently uses $\alpha = 0.3$, although $\alpha$'s best value has not been extensively tested. Moreover, a "best value" is likely highly dependent on the TOA data. \code{APTB} uses $\alpha = 0.3$, and not $\alpha = 1$ as in \code{APT}, because the scoring system of \code{APT} often over-weights particularly dense clusters that are not close enough to other clusters to allow phase connection to proceed. 

\subsection{Phase Wraps}
\label{sec:phasewraps}
When observing a pulsar, it is necessary to already have good estimates of the relevant parameters. Thus a single observation should correctly identify the relative pulse numbers of its TOAs. In other words, a cluster should be phase connected internally. In the case that a cluster is not phase connected internally, \code{APTB} can often correct for the TOAs that break phase connection by tracking the phase during the cluster.
The arbitrary definition of zero phase can cause the model to confuse residuals that cross the $\pm0.5$ rotation boundary. This makes these TOAs appear to lack phase connection when they otherwise would be phase connected. \code{APTB} corrects these instances by applying appropriate phase wraps on individual TOAs until phase connection within every cluster holds. Applying a phase wrap changes the assigned phase of a single or group of TOAs by an integer step. The correction is done quickly via the {\tt NumPy} {\tt unwrap} routine.

The initial timing ephemeris will usually not provide phase connection within every individual cluster nor between most of the clusters. Once every cluster has been corrected to be phase-connected within itself, it is necessary to manually change the assigned pulse numbers for each cluster of TOAs so that at least the relative rotation count between two clusters can be compared. 
The goal of phase connection is to correct the relative rotation count between each and every cluster.
The process of manually changing the pulse numbers of an entire cluster from those predicted by the (yet-to-be-correct) model is known as \emph{checking for phase wraps}. See Table \ref{tab:phase_table} for an example. 
Once phase wraps are used on individual clusters to phase connect the clusters within themselves,
phase wraps are then only evaluated between JUMPed, or individually phase-connected, sections of the TOAs in order to map a gap (see Section \ref{sec:gapmap}).

\begin{table}
	\centering
 	\begin{tabular}{ cccc } 
		\hline
		Cluster 6$_{M}$ & Cluster 7$_{M}$ & Cluster 6$_{C}$ & Cluster 7$_{C}$\\
		\hline
		2402 & 3534 & 2301 & 3432\\
		2607 & 3804 & 2506 & 3702\\
	    2803 & -- & 2702 & --\\
		\hline
	\end{tabular}
	\caption{
	Example of the global rotation numbers for the TOAs in Cluster 6 and Cluster 7 
	respectively. Columns 1 \& 2 are the model, M, predictions, and Columns 3 \& 4 are the correct, C, predictions. Cluster 6 has 3 TOAs while Cluster 7 only has 2. 
	While the model predicts the wrong 
	pulse numbers for each TOA, the difference in pulse numbers ($\Delta PN$) within each cluster \emph{are} correct 
	(i.e. $2506-2301=205=2607-2402$). 
	The $\Delta PN$ between Cluster 6$_M$ and Cluster 7$_M$ is 731, while the $\Delta PN$ between Cluster 6$_C$ and Cluster 7$_C$ is 730. Thus, Cluster 6 for this model needs a phase wrap of -1 to be corrected.}
	\label{tab:phase_table}
\end{table}

\subsection{Mapping a Gap}
\label{sec:gapmap}
Checking for phase wraps between clusters can be a computationally expensive process. It can be unclear how many possible phase wraps should be checked. Furthermore, if too many phase wraps are checked, the total number checked for every neighboring pair of TOAs grows quickly. See Section \ref{sec:phase_wrap_search_space} for a more detailed explanation of what we call the phase wrap search space, but in short, if an algorithm is not careful, it may be in the process of checking more than $10^{24}$ phase wraps, and even at a fraction of a second per phase wrap, this would take on the order of $10^{15}$\,years. 

Table \ref{tab:phase_table} shows an example where a cluster needs a single phase wrap removed to be correct. Unfortunately, we do not know this \emph{a priori}. To attempt to determine if phase wraps are needed, \code{APTB} uses a technique referred to as \emph{mapping a gap} (FR18). 
Below is a description of the gap-mapping technique used by \code{APTB}

Let $\chi^2_\nu (n)$ be the $\chi^2_\nu$ after applying a phase wrap of $n$. If a phase wrap of $m$ minimizes $\chi^2_\nu (n)$, then a Taylor series of $\chi^2_\nu (n)$ centered around $m$ has no linear $n$ term and so the $n^2$ term is the leading non-constant term. Therefore, the $\chi^2_\nu (n)$ has a quadratic dependence on the phase wrap when $n$ is close to $m$ (see also Figure 5 of FR18). 
Finding the vertex in phase wrap-$\chi^2_\nu (n)$ space gives $m$, which represents the best phase wrap according to the best, but not necessarily fully correct, model. The parabola can be defined by sampling the $\chi^2_\nu (n)$ for three different phase wraps (i.e. by sampling three phase wrap-$\chi^2_\nu(n)$ points) and the phase wrap of the vertex is given by the closest integer (i.e.~$\mathtt{round}[m]$) to
\begin{equation} 
    m=\frac{b}{2}\frac{\chi^2_\nu(-b) - \chi^2_\nu (b)}{\chi^2_\nu(b) + \chi^2_\nu(-b) - 2\chi^2_\nu(0)}.
	\label{eq:quadratic}
\end{equation}
This is the formula for calculating the vertex x-coordinate of the parabola $f(x) = a_2x^2 + a_1x +a_0$ by sampling the values of $f(b)$, $f(-b)$, and $f(0)$. \code{APTB} uses $b=5$, as do FR18, though this can lead to an error if $b=5$ gives a nonphysical parameter value, like a negative value of the projected semi-major axis ($a\sin i$). If this occurs, $b=4$ is attempted, repeating lower $b$ values until either the vertex is successfully calculated or $b=0$. If the $b=0$ iteration is reached, we cannot solve for the vertex of the parabola and the model is likely too poor to successfully phase connect, and so \code{APTB} will exit. A maximum $b=5$ is chosen because $b \lesssim 3$ would let small variations in the $\chi^2_\nu$, such as from TOA noise, greatly affect $m$ in Equation~\ref{eq:quadratic}. Additionally, $b \gtrsim 7$ would likely involve sampling phase wraps far from the phase wrap minimum---as the minimum is commonly at a phase wrap of order unity---weakening the quadratic-dependence assumption.

As for \texttt{DRACULA}, the gap-mapping method can only be successful if the TOAs are accurate to within their predicted error. If even a few TOAs are entirely erroneous, \code{APTB} will be fitting the parameters to compensate for fictitious differences in phase. Even worse, incorrect phase wraps will be treated as correct, and any model stemming from a model with even a single incorrect phase wrap can never be entirely phase connected when additional correct TOAs are included. 
Before phase connection is complete, it is nearly impossible to determine outliers because the phase connection process is sensitive enough to require the assumption of correct data.
Therefore, it is prudent that \code{APTB} is given correct TOAs that are as accurate as possible, with special attention given to removing TOAs with high predicted error. 
APTB cannot currently address the cases where some TOAs are in error. A potential improvement to the algorithm would be to have branches (Section \ref{sec:tree}) investigate the removal of TOAs or entire clusters.

\section{Solution Tree}
\label{sec:tree}

We begin this discussion by comparing pulsar timing to exploring a tree-like structure. The root or trunk of the tree is the starting model and different phase wrap decisions form branches. As different phase wraps are applied to attempt phase connection, more branches are created. This can lead to an exponential time complexity ($\mathcal{O}(2^n)$ where $n$ is the number of clusters) if the algorithm accepts too many possible phase wrap options (Section \ref{sec:phase_wrap_search_space}). Our approach assumes that several incorrect branches give a large $\chi_\nu^2$ early on in the phase-connection process so that these branches can be quickly discarded.

The best phase wrap in the short term is not necessarily the best phase wrap in the long term. In many cases, several possible phase wraps in a gap can give a $\chi^2_\nu$ that passes the pruning condition (see Section \ref{sec:pruning}). The $\chi^2_\nu$ for adjacent phase wraps can also be very similar, so even deciding the ``best'' phase wrap can be somewhat arbitrary, or at least weakly justified. Furthermore, we know our best-guess model is incorrect on some level so there is no reason to only choose the lowest $\chi^2_\nu$ phase wrap when this $\chi^2_\nu$ is based on a not-yet correct model. It thus becomes necessary to explore all acceptable branches, where acceptable is defined as having a $\chi^2_\nu$ below a predetermined value known as the pruning condition. We used the Python package \code{treelib}\footnote{\url{https://pypi.org/project/treelib/}} to manage traversal and pruning of the tree and its requisite bookkeeping, though other Python data tree libraries could serve the same purpose.

\subsection{Branching}
\label{sec:branch}

\begin{figure}
	\includegraphics[width=\columnwidth]{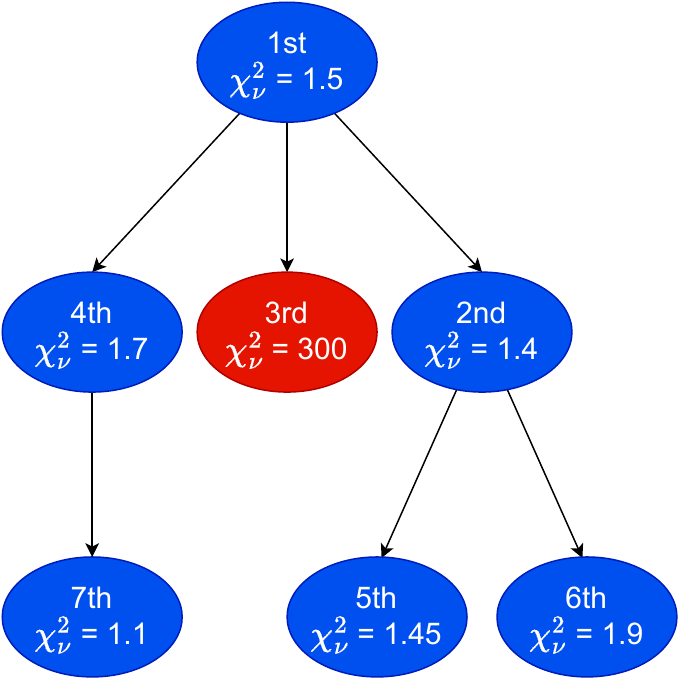}
    \caption{Flowchart of \code{APTB}'s branch searching priority. Even though the bottom left model has the lowest $\chi^2_\nu$, its parent prevented it from being immediately explored. Branching is based on phase wrap decisions so the 2nd, 3rd, and 4th models represent different phase wrap decisions that phase connect one more cluster than the 1st model. For example, the 2nd model could result from a phase wrap of 0, the 3rd model from a phase wrap of 1, and the 4th model from a phase wrap of -1. 
    Keep in mind no model is visible until it is attempted, so the 3rd and 4th models had to be attempted to know the 2nd model was the best. The red model was pruned so none of its children are explored. }
    \label{fig:solution_flow}
\end{figure}

Going down all acceptable branches is like trying to produce a tree-like structure that is actively forming with every decision made. The tree's structure is only revealed by exploring it. \code{APTB} creates the tree structure by mapping each gap and testing phase wraps. 
Phase wraps that fail the pruning condition result in that branch being cut. Phase wraps that pass are ranked based on lowest to highest $\chi^2_\nu$. The best phase wrap (lowest $\chi^2_\nu$) in the short term is explored first, with its own branches being explored next. Thus, this is a depth-first search (see Figure \ref{fig:solution_flow}).

\subsection{Pruning}
\label{sec:pruning}
While exploring several branches is often needed to phase-connect pulsars with limited data, it is important to know when a dead end is reached. In pulsar timing, there are an infinite number of options available so we have to be clever in deciding which branches, or phase wraps, are in fact dead ends. Thus a branch must be pruned when it becomes clear that it cannot lead to the correct solution.

When it is time for a cluster to be unJUMPed, the gap-mapping technique is applied. Again, gap-mapping allows \code{APTB} to know a reasonably good phase wrap to start with. The $\chi^2_\nu$ is calculated for the vertex point as well as for the phase wraps of $\pm$1 from the vertex. The minimum $\chi^2_\nu$ of these three phase wraps, now referred to as $PW_{\text{calc}}$, must be lower than the pruning condition. The pruning condition, $PC$, used by \code{APTB} is $PC = \chi^2_{\nu,\text{ base}} + 1$. If $PW_{\text{calc}} \geq PC$, then the F-test for model comparison is done on the current model vs. the current model plus a new parameter. The F-test provides a way to compare two models that have different degrees of freedom. We refer to Section 3 of PR22 for a brief discussion on the F-test for model comparison. This model is given one more chance and the gap-mapping technique is applied one more time, with $PW_{\text{calc}}$ calculated again. If $PW_{\text{calc}} \geq PC$ still, this branch is pruned. This model is given a second chance because in some instances evaluating the F-test with different parameters should be done before mapping the gap.

When a branch is created, certain parameters are checked to see if they are physically relevant. 
For instance, if $a \sin{i}$ (the projected semi-major axis of the orbit) is negative, then the branch is pruned. 
Optionally, a branch is pruned if $\dot f$ ($F1$) becomes positive, though this is not recommended as $F1$ could become negative in a later model stemming from this branch. However, a pulsar can be observed to have a non-intrinsic positive $F1$ if it has a negative radial acceleration. This is common for pulsars in globular clusters due to the gravitational potential of the clusters. In fact, two of the pulsars discussed in Section \ref{sec:results}, PSRs~J1748$-$2446aq and J1748$-$2446at, have a positive $F1$.
Also optionally, if the right ascension (R.A.) or declination (decl.) leaves a set boundary, the branch is pruned.
No other pruning mechanisms are implemented currently.

\subsection{Phase Wrap Search Space} \label{sec:phase_wrap_search_space} 

We pause the explanation of the algorithm to elaborate on why and when \code{APTB} may succeed or fail.
The models that \code{APTB} explores based on its different phase wrap decisions all comprise the phase wrap search space (PWSS). The size of this space can be the dominant factor in determining whether \code{APTB} will succeed or not. In the case of Figure \ref{fig:solution_flow}, the size of the PWSS is 7 models. Generally speaking, the size of the PWSS, $S$, is 
\begin{equation}
    S = \sum _{d = 0} ^ {N_{\text{clusters}}} W(d) ,
\end{equation}
where $N_{\text{clusters}}$ is the number of clusters and $W(d)$ is the number of models, or width, of the solution tree at a depth $d$. The depth refers to the number of clusters already phase connected for a given model minus one. In Figure \ref{fig:solution_flow}, $W(0) = 1$ and $W(1) = W(2) = 3$. If \code{APTB} takes $\tau$ seconds on average per model, then the total run time would be $\tau S$. 

The nature of $W(d)$ marks the boundary between a solvable and unsolvable pulsar using our algorithm. We will start with just one example where \code{APTB} would fail. If every gap has two plausible phase wraps (two branches), then $W(d) = 2^d$, implying $S \geq 2^d$. For the case of PSR~J1748$-$2446aq (Section \ref{sec:results}), $\tau \sim 10$\,s and $N_{\text{clusters}} \sim 80$. Therefore, the total run time would be of order $10\cdot2^{80} \text{\,s} \sim 10^{17}$\,yr. Clearly, APTB would be unsuccessful for PSR~J1748$-$2446aq if it needed to check two branches for every gap. 

\code{APTB} can be successful due to most pulsars following a different $W(d)$. The initial model is precise enough to phase-connect TOAs within an observation but not between observations (i.e.~clusters). Therefore, $W(1)$ can be quite high, and $W(2)$ through roughly $W(5)$ can become exponentially higher. Eventually, a model has phase-connected enough clusters to be much more precise in its predictive power and so admits exponentially fewer possible branches, significantly lowering the width at higher depths. 
This all suggests we adopt
\begin{equation}
    W(d) \sim 
    \left\{
        \begin{array}{lr}
             r_1^d, & \text{if } d \leq d_0\\
             r_1^{d_0} r_2^{(d-d_0)}, & \text{if } d > d_0\\
        \end{array}
    \right\},
\end{equation}
where $r_1$ and $r_2$ are the growth and decay rates, respectively, and dictate $W(d)$ based on the cutoff depth, $d_0$. 
The values of $r_1$, $r_2$, and $d_0$ will determine if \code{APTB} can solve a pulsar. Non-integer values of $W(d)$ can be interpreted as a characteristic value. Assuming $W(d \lesssim d_0) \gg W(d \gg d_0)$, we can simplify further to $S \sim \sum _{d = 0} ^ {d_0} W(d) \sim r_1^{d_0}$. Therefore if either $r_1$ or $d_0$ is too large, $S$ will also be too large. $r_1$ represents how quickly the model will branch until a depth of $d_0$ is reached. Pulsars with bad data (e.g.~due to erroneous TOAs or low cluster density) will initially accept many phase wraps (increasing $r_1$) and could continue to accept them for longer (increasing $d_0$).

While both human and computer-based timing rely on the search space not becoming too large, a computer is able to probe a much larger search space. This is helpful in the cases where the cutoff depth allows for $S$ no larger than on the order of hundreds of models to a few thousand models, but not when the search space balloons to tens of thousands of models or more. A human could take days to search $S$$\sim$100 models but our algorithm can take minutes or hours and is fully automated. This is not to mention that \code{APTB} would be more systematic than a human, searching for phase wraps in the same way every time. When the search space is in the tens of models, then both humans and \code{APTB} would find no issue, but \code{APTB} would still likely phase connect faster.

This is all to say that while \code{APTB} can solve trickier pulsars than a human could solve, \code{APTB} can still fail even when functioning as intended. Thinking of our algorithm with the PWSS in mind can give a first-level analysis of the limits of our algorithm.



\section{Binary Models}
\label{sec:bmodels}

It is important to choose the appropriate binary model as each model has its own unique benefits in terms of not only enabling the possibility of phase connection but also the overall computational efficiency of phase connection. Thus, the binary model should already be selected by the user, with proper estimates of the binary parameters established. Currently, \code{APTB} can handle the ELL1 \citep{lange2001precision}, BT \citep{blandford1976arrival}, and DD \citep{damour1985general, damour1986general} models.  
Implementing additional binary models would be straightforward.

\subsection{Nearly Circular Orbits}

Most binary MSPs have nearly circular orbits where is it difficult to measure the longitude of periastron ($\omega$) and epoch of periastron passage ($T_0$). 
This is because the TOAs do not clearly indicate the location of periastron, and thus the epoch of periastron is not clearly indicated either \citep{lange2001precision}.  
The ELL1 model was designed for nearly circular orbits to address this difficulty by measuring the orbital phase with respect to the epoch of ascending node ($T_{\text{asc}}$) rather than $T_0$. 
The ELL1 model is a Newtonian binary model that neglects eccentricity ($e$) terms of order $e^2$ or higher, although most timing software have improved the accuracy of the ELL1 model by adding $e^2$ or even $e^3$ terms. Therefore, this model describes residuals that follow a slightly perturbed sinusoid as a function of the orbital phase. 
Part of \code{APTB}'s initialization is to not start fitting for $\epsilon_1 = e \sin{\omega}$ and $\epsilon_2 = e\cos{\omega}$ immediately. $\epsilon_1$ and $\epsilon_2$ can only be fit after the unJUMPed clusters span 5 times the binary orbital period ($P_b$), though this can be changed by the user. In contrast, the model should immediately begin fitting for $P_b$, $T_{\text{asc}}$, and $a\sin{i}$ in order to correct for orbital effects within every cluster. Also, for this reason, these parameters should be known to a satisfactory precision beforehand. Notably, when the F-test processes are conducted (Section \ref{sec:pruning}), both $\epsilon_1$ and $\epsilon_2$ are tested together. If they significantly help the model they are added to the model together. In many cases, the eccentricity is negligible so assuming $\epsilon_1 = \epsilon_2 = 0$ is often sufficient for phase connection.

It should be noted that while using the ELL1 model, \code{APTB} does admit post-Keplerian parameters such as the time derivative of the binary period ($\dot{P_b}$). These are very rarely required for phase connection on short timescales, though, and are usually included after phase connection has already been achieved. We briefly discuss the use of $\dot{P_b}$ in the timing of PSR~J1748$-$2446ar in Section \ref{sec:results}. Post-Keplerian parameters are only handled if the user explicitly includes them in the initial parameter file.

\subsection{Non-Circular Orbits}
Though less common, some binary pulsars are elliptical enough to require binary models that make no assumptions about the eccentricity.
One such model is the BT model, a Newtonian model. 
If the eccentricity is important for phase connection purposes, this binary model should be used. \code{APTB} will begin fitting for all major binary parameters immediately, and therefore no testing via the F-test is done on these parameters. The base parameters of the BT model are $P_b$, $a\sin{i}$, $T_0$, $e$, and $\omega$. Like in the ELL1 model, \code{APTB} can also handle post-Keplerian parameters provided the user includes them in the initial parameter file.


Another non-circular orbit model is the DD model, a post-Newtonian approximation relativistic model that is theory-independent. This model is intended for pulsars with large eccentricities and measurable secular and periodic post-Newtonian effects. Similar to the BT model, all major parameters will be fit for immediately. These parameters are $P_b$, $a\sin{i}$, $T_0$, $e$, $\omega$, and the time derivative of the longitude of periastron ($\dot{\omega}$).

\section{Algorithm}
\label{sec:algorithm}

\begin{figure*}
        \centering
	\includegraphics[width=\textwidth]{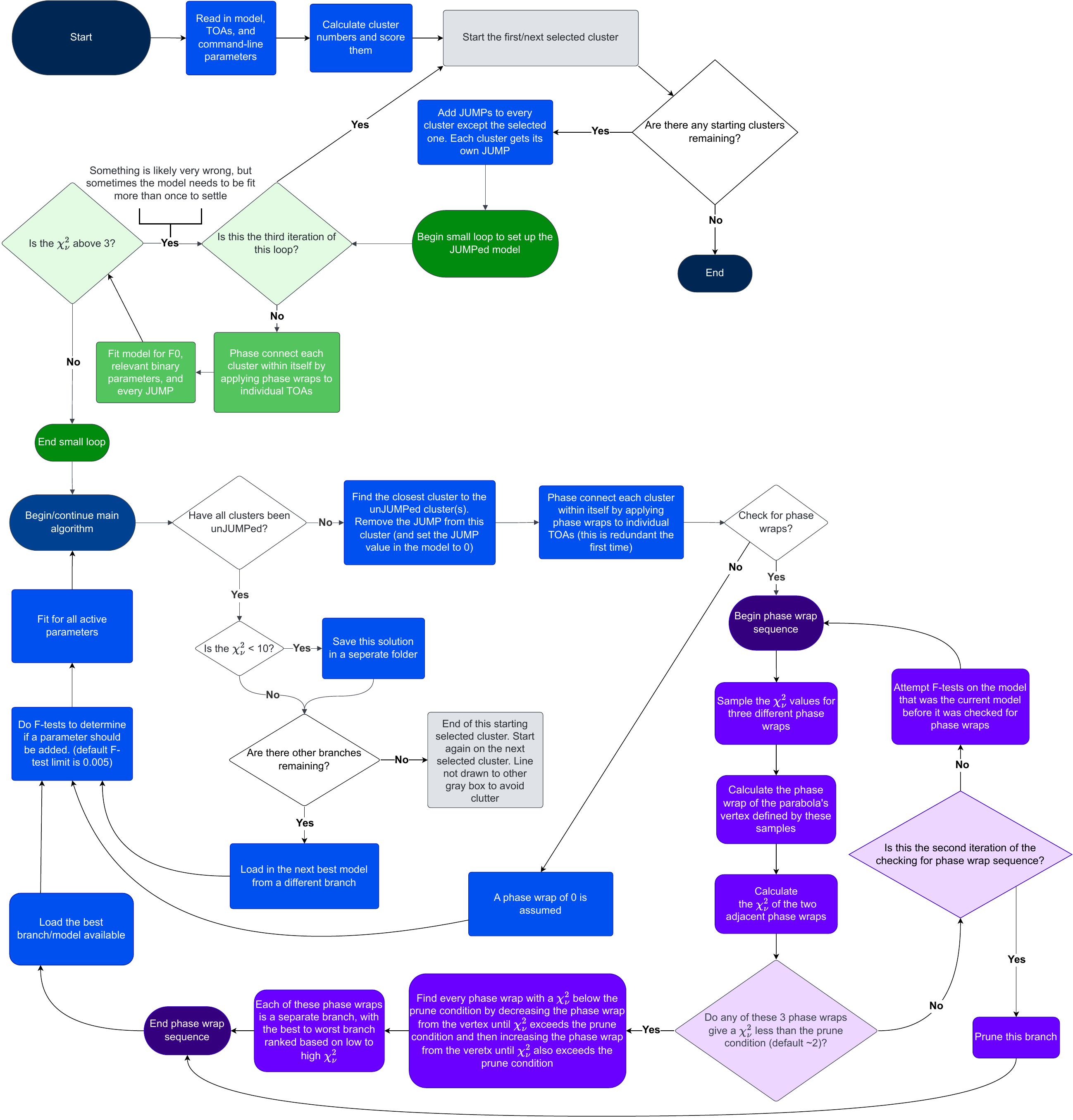} 
    \caption{Flowchart of \code{APTB}'s logic. }
    \label{fig:APTB_flowchart}
\end{figure*}

\code{APTB} is an algorithm that uses JUMPs, the F-test, gap-mapping, and a data tree structure in order to phase connect isolated and, more specifically, binary pulsars.
We strongly recommend using the flowchart given in Figure \ref{fig:APTB_flowchart} to help guide the rest of this section. 
Before diving into the specifics of the algorithm, it is helpful to describe it on a higher level. The best starting cluster is selected and \code{APTB} will attempt to remove JUMPs from more and more clusters to eventually phase connect every TOA. To do this, adjacent clusters to the starting cluster are unJUMPed. We know the model is incorrect on some level, yet should give almost the correct relative pulse number difference between clusters. To verify this, we determine the $\chi_\nu^2$ of various phase wraps. This should have a parabolic dependence assuming the model is not too incorrect. Then the most plausible phase wraps can be investigated. Using a depth-first search technique, \code{APTB} will investigate all branches.
If the initial starting cluster yields no acceptable solutions, \code{APTB} will try other starting clusters. 

\subsection{Algorithm In Depth}
\label{sec:algorithm_detailed}

The model is initialized by applying JUMPs to every cluster except one. \code{APTB} then ensures that each cluster is phase connected internally. The highest-scoring cluster, as defined by Equation~\ref{eq:score}, is chosen as the starting cluster and is the only cluster not JUMPed. If the $\chi_\nu^2$ is above three at this point, there may be a problem that \code{APTB} cannot fix. \code{APTB} will attempt to fix this by phase connecting and fitting two more times, but this is likely to fail. Either the starting timing model is too poor, the TOAs are inaccurate, or the TOA uncertainties are underestimated. In any case, the user must do something on their end before proceeding. This can include fixing one of the above problems or overriding this warning (see Section \ref{sec:options}).

Provided the model has been properly initialized, the main algorithm can begin. The starting cluster is the first member of the \emph{unJUMPed clusters}. The cluster closest in time to the unJUMPed clusters is unJUMPed and therefore added to this group. The number of pulsar rotations between the closest cluster and the other members is not known, so the phase wraps between the closest cluster and the previously phase-connected data are checked according to the gap-mapping methodology (Section \ref{sec:gapmap}). Gap-mapping (ideally) finds the minimum $\chi_\nu^2$ for any particular phase wrap, but another phase wrap may indeed prove to be the correct one. For this reason, \code{APTB} incrementally checks the phase wraps $1, 2, 3, ..., N_\text{below}$ below the best phase wrap until the model fails the pruning condition at a phase wrap of $N_\text{below}+1$ below the best phase wrap. The same is repeated for $1, 2, 3, ..., N_\text{above}$ above the best phase wrap, with the model failing the pruning condition at the $N_\text{above}+1$ phase wrap above the best phase wrap. The phase wraps between $N_\text{below}$ and $N_\text{above}$, inclusive, are acceptable models that represent branches on the solution tree.

All acceptable solutions are stored at this point, and attached to the parent model that created them. The next used model is determined by the branch search hierarchy (see Section \ref{sec:branch} and Figure \ref{fig:solution_flow}). 
This next model is then fit for new parameters if they significantly improve the model. As in PR22, an F-test value of 0.005 or lower is deemed a significant improvement (i.e. the probability that the model fit improvement could be due to noise is $<$0.005). Not every non-fit parameter is tested. 
The right ascension and declination can be tested after the unJUMPed clusters span at least $T_{\text{RA}}$ and $1.3 \cdot T_{\text{RA}}$, respectively, where
\begin{equation}
    T_{\text{RA}} = \left(-\frac{30}{700}\cdot \frac{f}{\text{Hz}} + 40\right)\,\text{days}.
    \label{eq:tra}
\end{equation}
The functional form of $T_{\text{RA}}$ is such that $T_{\text{RA}}$ depends linearly on the pulsar's spin frequency, $T_{\text{RA}}(f=700 \text{\,Hz})$ = 10\,days, and the maximum cluster span until the right ascension is tested is 40\,days (i.e.~$\mathtt{max}[T_{\text{RA}}]$ = 40\,days). 
The linear dependence condition is motivated by the fact that the sky position becomes important quickly for pulsars with higher TOA precision, which is usually directly proportional to the spin frequency.
The three requirements for Equation \ref{eq:tra} have not been derived from first principles, but rather from experience phase connecting many pulsars manually. 
$F1$ can be tested after the unJUMPed clusters span is long enough that a typical $F1$ would cause the residuals to exceed $\pm$0.35 in phase. A typical $F1$ is estimated based on the pulsar's spin frequency and rough placement in the $P$-$\dot P$ diagram (see e.g.~Figure 6.3 of \citet{condon2016essential}). 

Currently, \code{APTB} has the capability to conduct an F-test and fit for $\ddot f$ ($F2$), but the user has to specify a minimum unJUMPed cluster time span since the default is not to test for $F2$. We choose to omit $F2$ by default because $F2$ usually only becomes significant in time spans longer than are necessary for phase connection. Furthermore, typical $F2$ values vary widely, and estimating when $F2$ should be fit is difficult; a wrong guess may prevent the correct solution from being found due to possible covariances with the other parameters. In most cases, $F2$ can be ignored until after a successful phase connection, and then manually fit, as we did for PSR~J1748$-$2446aq (Section \ref{sec:results}).


After the appropriate F-tests and new parameter additions, the model is then brought back to the beginning of the algorithm, as all model selection from the solution tree is done after checking for phase wraps. The first step is, again, to unJUMP the closest unJUMPed cluster, and so on.

Once every TOA has been unJUMPed, if $\chi_\nu^2 < 10$ then the model is deemed a possible solution and saved. \code{APTB} then moves to the next starting cluster to see if another (or the same) solution arises.

In some cases, every branch is pruned before finding a solution. This means \code{APTB} has failed to find the solution with this particular starting cluster. The algorithm will attempt to run again with the second-best starting cluster and will keep selecting starting clusters until the top five (default) clusters by score have been attempted. While we hope to make \code{APTB} robust enough to find the solution regardless of a starting point, some assumptions break down when the model is too poor or the data is too sparse. When searching for correct phase wraps, \code{APTB} uses the quadratic dependence of the $\chi_\nu^2$ to efficiently find the minimum. If the minimum phase wrap is too large or the model is too poor, this quadratic dependence falls apart quickly. This could pin every early-on model into a local minimum rather than the global minimum. If this local minimum is too wide, as in the parabola covers a large number of phase wraps, or if there are too many local minima, then \code{APTB} may not find any solutions or a very large number of wrong solutions.

\code{APTB} will store all explored models on the user's local disk. These files are crucial in the case \code{APTB} is very close to finding a solution but failed due to unforeseen circumstances. For example, if the model requires $F2$ or the derivative of the binary orbital period, it may have pruned the correct branch. By finding this almost-complete model, the remaining few steps can be manually completed quickly. For more details on the files saved and how to use them, see the \code{APTB} User Guide.

\subsection{Optional Features}
\label{sec:options}

\code{APTB} comes equipped with several ways to modify its behavior from the command line. We will describe the most useful modifications, with the rest being listed and briefly described in the \code{APTB} User Guide and the Python file on GitHub. By default, \code{APTB} will check for phase wraps and also use the solution tree structure. These can separately be turned off by including 
{\tt -{}-no-check\_phase\_wraps} and {\tt -{}-no-branches}, respectively,  in the command line.  But note that branching only works if phase wrap checking is on. If phase wrap checking is turned off, then \code{APTB} will assume that the cluster that just had its JUMP removed has a phase wrap of zero with the unJUMPed clusters. In other words, at each step, \code{APTB} assumes the model correctly predicts the relative pulse numbers after removing a JUMP.

If phase wrap checking is on but branching is not, then \code{APTB} will map the gap and find the phase wrap with the lowest $\chi_\nu^2$, but will not pursue nearby phase wraps that are also acceptable. For some pulsars, this is sufficient to time them successfully.
While \code{APTB} is really meant to excel in the case where both phase wrap checking and branching are \emph{required} to time the pulsar, the algorithm can also easily and quickly time the less challenging case. 

When \code{APTB} finds a potential solution, there may be other unexplored branches in the solution tree that may yield other potential solutions.  
By default, \code{APTB} will explore these branches to rule out, or discover, any degenerate solutions.
This can be turned off, but doing so can have similar effects to turning off branching in that the correct solution may not be initially the best one. The advantage to stopping \code{APTB} from pursuing branches after finding a potential solution is that there may be a large number of potential solutions (tens or hundreds) which can consume much more computing time and storage than anticipated. 

Some command line options affect the pruning condition. By default, if $\chi^2_{\nu,\text{ base}} > 3$ then the algorithm will not proceed. This can be overridden if the user thinks \code{APTB} should try to time this pulsar anyway.
The pruning condition by default is set to $\chi^2_{\nu,\text{ base}} + 1$ but the user may specify their own value. 
When \code{APTB} can fit for R.A., decl., $F1$, and certain binary parameters like $\epsilon_1$ and $\epsilon_2$ can be adjusted manually. 
Lastly, a rectangle in R.A.-decl. space can be specified. A model will be pruned if its astrometric position leaves this rectangle.

\section{Results} \label{sec:results}

One important test of \code{APTB} is whether it can time and solve a pulsar that has never been phase-connected before.
Indeed, we demonstrate the capability of \code{APTB} by showing how it was first in determining the phase-connected timing solutions for the newly discovered PSRs~J1748$-$2446aq and J1748$-$2446at. We also show that it was able to verify phase-connected timing solutions for the PSRs~J1824$-$2452N and J1748$-$2446ar.

\begin{figure}
    \centering
    \includegraphics[width=\columnwidth]{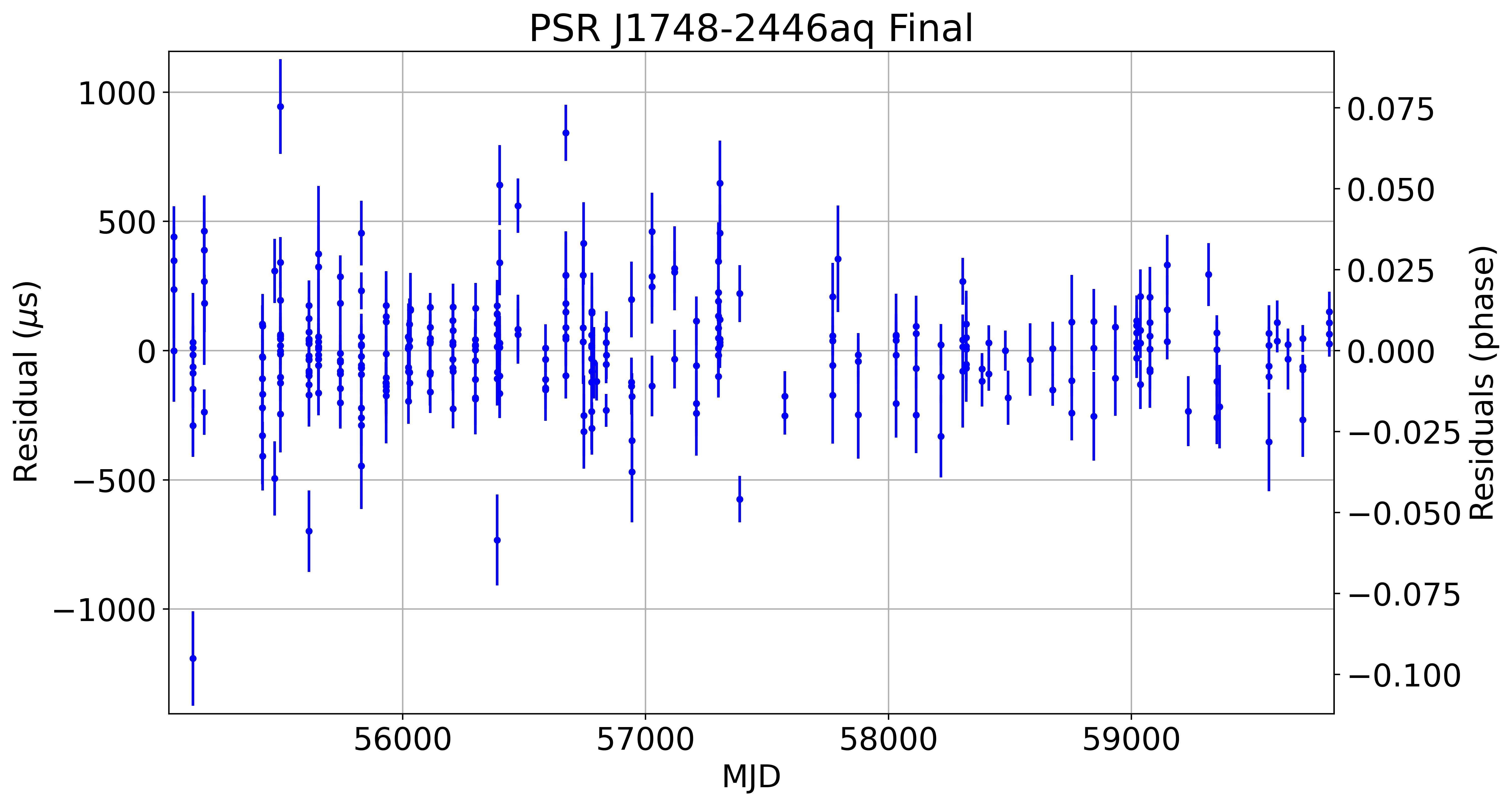}
    \caption{TOA residuals for PSR~J1748$-$2446aq after \code{APTB}'s successful phase connection.}
    \label{fig:psrter5aq_final}
\end{figure}

PSR~J1748$-$2446aq, Ter5aq hereafter, was discovered in February of 2023 (Padmanabh et al., in preparation) and is the forty-second pulsar discovered in Terzan 5.\footnote{\url{https://www3.mpifr-bonn.mpg.de/staff/pfreire/GCpsr.html}}
Terzan~5, a globular cluster near the galactic bulge, is home to the largest globular cluster pulsar population \citep{martsen2022radio}, with several recent new discoveries. 
Ter5aq has 86 clusters available spanning 13 years, which gives an average cluster density of roughly 0.018 clusters/day or 1 observation every 55 days. Most pulsars have a much higher observation cadence. Moreover, no observations in the TOA data provided to \code{APTB} were dedicated specifically for phase connection purposes. That is, \code{APTB} did not require a grouping of clusters with a significantly higher cluster density than the average.
With \code{APTB} phase connecting Ter5aq first, this verifies that \code{APTB} can time low cluster density pulsars. 
We show the final residual plot in Figure \ref{fig:psrter5aq_final}.
Ter5aq is a millisecond pulsar with a 13\,ms spin period, and is in a binary system with an orbital period of 0.12\,days and a 0.026\,light-second projected semimajor axis.
The full solution from \code{APTB}, including additional data, will be published in Padmanabh et al. (in preparation).

\begin{figure*}
    \centering
    \centering
        \includegraphics[width=\textwidth]{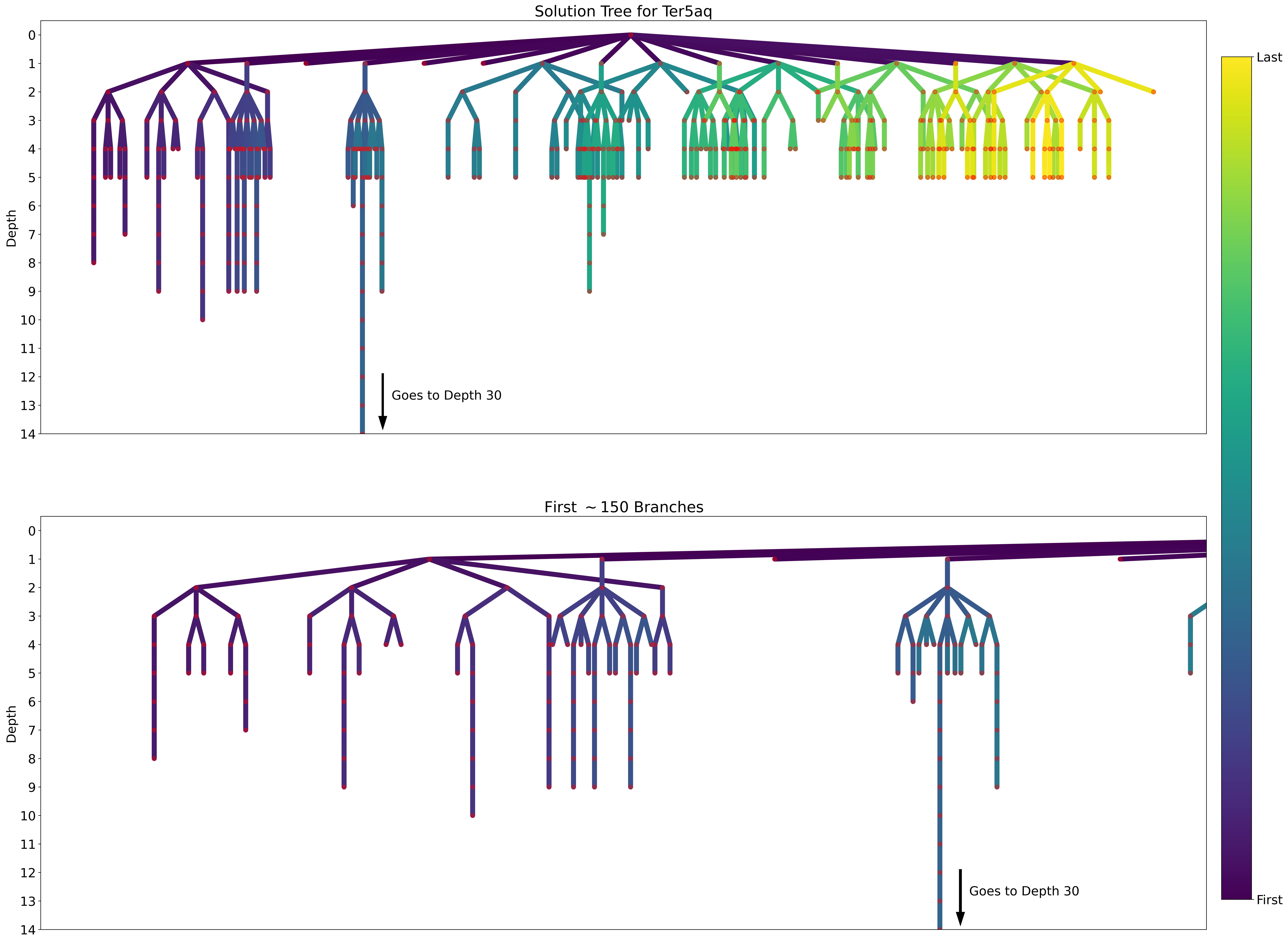}
        \label{fig:ter5aq_tree_full}
    \caption{A schematic of the solution tree \code{APTB} investigated in order to time Ter5aq. The branching order that \code{APTB} took is qualitatively described by the color bar. The dark purple lines show branches that \code{APTB} investigated first while the light green and yellow lines show branches \code{APTB} investigated last. The depth of the solution tree is equivalent to the number of clusters that had been unJUMPed at that stage minus one. The green and yellow branches were investigated heavily, but no meaningful solution was found. 
    The dark blue-turquoise branch that goes past a depth of 14 actually ends at a (not shown) depth of 30.
    This emphasizes that the depth-30 branch in fact leads to the unique solution. The lower panel shows only the first $\sim$150 branches investigated.}
    \label{fig:ter5aq_tree}
\end{figure*}

To reinforce the difficulty a human would have in timing Ter5aq with the available data, we include Figure \ref{fig:ter5aq_tree}, which schematizes the phase wrap decisions \code{APTB} made. \code{APTB} took 416 iterations to complete its search, and 106 iterations of the first 109 iterations were all on branches leading to a dead end. It was not until the 110th iteration that \code{APTB} was on the correct branch and stayed on it. 
The large phase wrap search space means \code{APTB} took 5.0 hours, or 1.0 hours on average per starting cluster. The first starting cluster that was successful took 1.5 hours. This may seem like a long run-time, but a human would likely take much longer and give up from frustration after many incorrect attempts.
Because $F2$ was excluded from being fit for, \code{APTB} only reached a depth of 30, instead of the full depth of 86 clusters. Therefore, \code{APTB} was not confident it found the correct solution, but upon inspection of the solution tree, where post-fit timing residuals are plotted and saved, it became obvious that the phase connection was near complete. Even after reaching a depth of 30, \code{APTB} continued to look for more solutions to ensure the solution was unambiguous. 
While Ter5aq was eventually timed independently by a human, it is easy to imagine a pulsar with even less dense observations or a smaller number of them. In this case, using the automatic capabilities of \code{APTB} might be the only option to phase-connect that pulsar. 


PSR~J1748$-$2446at, Ter5at hereafter, was discovered in April 2023 (Padmanabh et al., in prep) making it the forty-fifth Terzan~5 pulsar (PSR~J1748$-$2446ar, see below, was discovered before Ter5at).
Ter5at is a black widow system, a type of binary pulsar with a companion of only a few hundredths of a solar mass \citep{shahbaz2017properties}. 
While also eventually phase-connected by a human, \code{APTB} determined the proper phase connection first. 
The 2.2\,ms spin period is significantly faster than that of Ter5aq, and its orbital period and projected semimajor axis are 0.22\,days and 0.10\,light-second, respectively. Again, the full solution will be published in Padmanabh et al. (in preparation).
The cluster density is 0.016 clusters/day (an observation every 62 days), and \code{APTB}'s runtime was 5.3\,hr. The first starting cluster did not lead to a solution, but the other four did. The second starting cluster found the solution the quickest and was on the correct phase wrap branch immediately. 
As in the case of solving Ter5aq, \code{APTB} reached a tree depth where the residual effects of $F2$ became important, in which case \code{APTB} thought it reached a dead end. This was easily spotted in the solution tree, and the process of finishing the phase connection was straightforward.

\code{APTB} can also time pulsars with data spans shorter than 14 years, provided the cluster density is higher. This is exemplified by the successful timing of PSR~J1824$-$2452N, M28N hereafter. M28N was discovered by \citet{douglas2022two} and is the fourteenth pulsar discovered in the globular cluster M28. \citet{douglas2022two} describe M28N as a Black Widow MSP with a binary period of 4.76\,hr. 
\code{APTB} was given only about 3 years of TOAs with a cluster density of roughly 0.035 clusters/day (an observation every 29 days), roughly twice that of Ter5aq. As is the case for pulsars with high TOA densities, \code{APTB} struggled less to time M28N but was still on a dead-end branch for its first 5 iterations. This pulsar was fit for only the first 900 days because a longer timescale makes $F2$ non-negligible. 

\code{APTB} has been able to time several more real pulsars with known phase-connection, but we will only mention one more instructive example. PSR~J1748$-$2446ar (Ter5ar hereafter) is the forty-third pulsar discovered in Terzan 5 (Padmanabh et al., in prep). This is a redback pulsar, a type of pulsar characterized by having a stellar companion with a few tenths of a solar mass \citep{shahbaz2017properties}. 
A redback pulsar's wind often ablates its companion, creating intra-binary material that eclipses the pulsar's radio pulses every orbit. 
The material, as well as the non-degenerate nature of the companion star, can cause years-long timescale perturbations in the residuals. The specific patterns of these effects appear to be largely random, making modeling and predicting pulses far into the future very tedious or impossible. 
As such, the $\chi^2_{\nu,\text{ base}} \sim 8$ is rather high, and this was after removing TOAs that were clearly erroneous. 
The high $\chi^2_{\nu,\text{ base}}$ can be explained by the fact that a redback system's orbital variability can be significant over even a couple of years.
For this reason, \code{APTB} was only given roughly the first year of data on Ter5ar. The derivative of the orbital period ($\dot{P_b}$) can be important on this timescale, with $\dot{P_b}$ and $\ddot{P_b}$ both becoming crucial on any timescale exceeding a year due to the short binary period and small $a \sin{i}$. In spite of these difficulties, \code{APTB} found the solution, and the correct branch was the first branch it investigated. \code{APTB}'s success is likely thanks to the relatively high cluster density of 0.071 clusters/day (an observation every 14 days).
Owing to the nature of the binary system, the final $\chi^2_\nu = 10.1$, which is high but comparable to the $\chi^2_{\nu,\text{ base}}$. 
No other plausible solution was found, and in spite of the abnormally high $\chi^2_\nu$, the found solution properly phase connects the next year of data. 



\section{Conclusion}
\label{sec:conc}

We have created an algorithm, \code{APTB}, that can phase-connect isolated and binary pulsars. 
Due to \code{APTB}'s robust branch-searching structure, it was able to time and solve PSRs~J1748$-$2446aq and J1748$-$2446at before any human could. 
\code{APTB} also timed several other known pulsars, including PSRs~J1748$-$2446ar and J1824$-$2452N. A standard successful run takes on the order of tens of minutes to several hours. In cases where \code{APTB} is unsuccessful in phase connection, the process can take much longer as the phase wrap search space expands exponentially. 


Like other pulsar timing algorithms, \code{APTB} allows a computer to make the same decisions that a human would. 
Even in the case of hour-long run times, this is of no concern to the pulsar astronomer as \code{APTB} is fully automated.
Therefore, our algorithm stands as an example of how current and future astronomers can grapple with ever-enlarging data sets. 
Pulsars discovered in globular clusters benefit from abundant archival data that an automated algorithm like \code{APTB} can easily handle, saving manual work-hours. Moreoever, a pulsar timing algorithm can become a necessity to successfully phase connect a binary pulsar when its data is too sparse. The increase in pulsar discoveries without a corresponding increase in available telescope time will only make sparse-data binaries more prevalent. 
While there already exists pulsar timing algorithms that can time binary pulsars, like \texttt{DRACULA}, \code{APTB} has the important advantage of being fully automated and therefore is much more suitable for binaries with sparser data.
Lastly, \code{APTB} is currently maintained, open to improvements, and is written exclusively in the modern and popular Python coding language. Several improvements, including support for new binary models and more options for weighting clusters, among others, are planned. 

Binary pulsars are some of the most scientifically interesting pulsar systems, and phase connecting them is crucial for determining their scientific potential. Tools like \code{APTB} allow us to accomplish that with less telescope time and less human effort.

This work was made possible under the guise of the National Radio Astronomy Observatory Research Experience for Undergraduates program funded by the National Science Foundation (NSF) through NSF grant number 1852401. The National Radio Astronomy Observatory is a facility of the National Science Foundation operated under cooperative agreement by Associated Universities, Inc. 
The Green Bank Observatory is a facility of the National Science Foundation operated under cooperative agreement by Associated Universities, Inc.
SMR is a CIFAR Fellow and is supported by the NSF Physics Frontiers Center award 2020265.



\typeout{}
\bibliographystyle{aasjournal}
\bibliography{main}{}





\end{document}